\documentclass{article}

\usepackage{PRIMEarxiv}

\usepackage[utf8]{inputenc} % allow utf-8 input
\usepackage[T1]{fontenc}    % use 8-bit T1 fonts
\usepackage{hyperref}       % hyperlinks
\usepackage{url}            % simple URL typesetting
\usepackage{booktabs}       % professional-quality tables
\usepackage{amsfonts}       % blackboard math symbols
\usepackage{nicefrac}       % compact symbols for 1/2, etc.
\usepackage{microtype}      % microtypography
\usepackage{lipsum}
\usepackage{fancyhdr}       % header
\usepackage{graphicx}       % graphics
\graphicspath{{media/}}     % organize your images and other figures under media/ folder

\title{Beam dynamics study of the high-power electron beam irradiator using niobium-tin superconducting cavity
\thanks{\textit{\underline{Citation}}: 
\textbf{Authors. Title. Pages.... DOI:000000/11111.}} 
}

\author{
  O. Tanaka, Y. Honda, M. Yamamoto, T. Yamada, H. Sakai,\\ 
  High Energy Accelerator Research Organization, KEK \\
 305-0801 Oho, Tsukuba, Ibaraki, Japan\\
  \texttt{olga@post.kek.jp} \\
}

\begin{document}

\maketitle

\begin{abstract}
A compact accelerator design for irradiation purposes is being proposed at KEK. This design targets an energy of 10 MeV and a current of 50 mA. Current design includes a 100 kV thermionic DC electron gun with an RF grid, 1-cell normal-conducting buncher cavity, and Nb$_{3}$Sn superconducting cavities to accelerate the beam to the final energy of 10 MeV. The goal of the present beam dynamics study is the beam loss suppression (to the ppm level), since it results in a thermal load on the cavity. Then the beam performance at the accelerator exit should be confirmed. The main issue was to transport the beam without loss, since the initial electron energy (100 keV) is low, and the beam parameters are intricately correlated. In addition, the space charge effect is considerable. For this reason, simultaneous optimization of multiple parameters was necessary. Here we report optimization results and their effect on the design of the machine.\par
\end{abstract}

% keywords can be removed

\section{Introduction}
At the Compact Energy Recovery Linac (cERL) at KEK \cite{akemoto}, an irradiation beam line was constructed and successfully commissioned \cite{Morikawa:2019bza}. Following the recent trend in accelerator science to design a compact high-current irradiation-type accelerators \cite{dhuley}, \cite{ciovati} and based on the results of irradiation experiments at the cERL \cite{Sakai:2022hxt}, we aim to develop a 50 mA electron beam irradiation facility working at the total beam energy of 10 MeV.\par

High-current beam acceleration can be achieved by using a superconducting cavities (SCs). In recent years, niobium tin (Nb$_{3}$Sn) accelerating cavities have attracted attention as next-generation accelerating cavities to replace niobium ones \cite{porter:linac18-tupo055}. A higher transition temperature (18.3 K) of Nb$_{3}$Sn makes it possible to operate the beam using only a simple small refrigerator instead of the conventional large He one. That allows to design a compact facility. The conceptual design of the proposed machine with special emphasize on the Nb$_{3}$Sn SCs implementation is given in  \cite{sakai-IPAC23-THPM124}.\par

\begin{figure}[ht]
   \centering
   \includegraphics*[width=0.7\columnwidth]{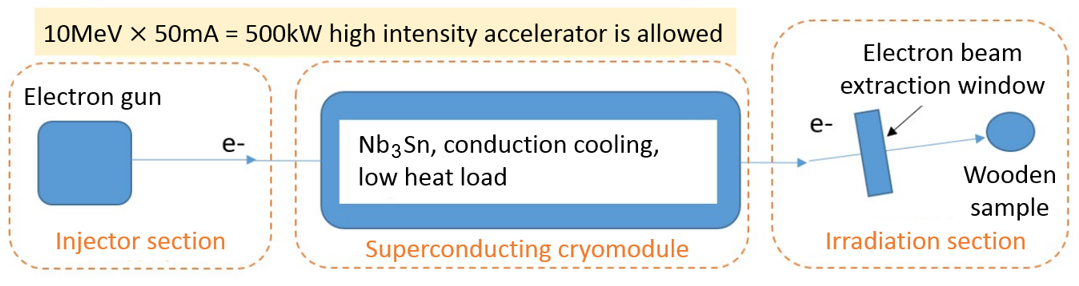}
   \caption{Conceptual design of 10 MeV, 50 mA accelerator using Nb$_{3}$Sn SRF cavities.}
   \label{fig:fig1}
\end{figure}

The overall design of the machine assumes the electron beam to be of 10 MeV and 50 mA. Fig.\ref{fig:fig1} shows a schematic of the entire system. It includes: an injector part with an electron gun of specific design, superconducting cryomodule, and an irradiation part, so that the machine can irradiate a large current beam. In order to evaluate the acceleration acceptance of an electron beam irradiation facility, we performed a three-dimensional particle tracking including the space-charge effect with the acceleration cavity system consisted of a buncher, two 1-cell SCs, and five 2-cell SCs. We found solutions to mitigate the beam loss applying a multi objective optimization. In the present study the beam dynamics issues associated with the irradiator design are discussed in details.\par

\section{System overview}
\label{sec:sys}
\begin{figure*}[ht]
    \centering
    \includegraphics*[width=0.7\textwidth]{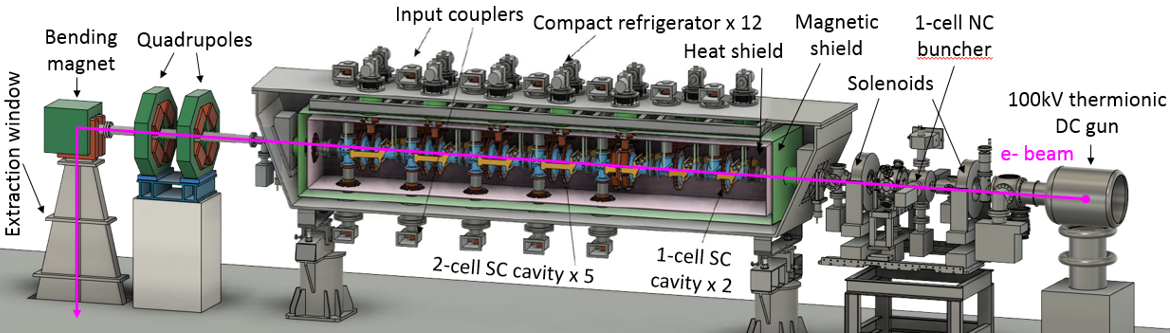}
    \caption{Layout of accelerator components.}
    \label{fig:fig2}
\end{figure*}

The electron gun is a 100 kV thermionic DC  gun with an RF grid that produces a repetitive longitudinally packed pulsed electron beam at 650 MHz \cite{bakker}.
After that, we assumed to compress the pulsed beam in the beam direction in a 1-cell normal-conducting buncher cavity, and then accelerate it to 10 MeV using two 1-cell SCs ($\beta=0.8$) and five 2-cell SCs. Note, low-$\beta$ resonators are just cavities that accelerate efficiently particles with velocity $\beta<1$ \cite{Facco:2004bb}. The RF frequency of all SCs is 1.3 GHz. The buncher cavity operates at 650 MHz. Two solenoids were installed before and after the buncher to provide transverse focusing. In the irradiation section there are two quadrupoles placed 1 m downstream the cryomodule to adjust the beam size. Then the beam is bent downward by the dipole and reaches the extraction window (See Fig.\ref{fig:fig2}).\par

The design of the electron gun system and the injection part was based on that of cERL \cite{akemoto}. A thermionic gun was selected for the electron source because it can stably supply a large current of 10 mA or more and an acceleration voltage of 100 kV \cite{fong}, as shown in Fig.\ref{fig:fig2}. A collimator was introduced in the injector design to limit the formation of a vacuum pressure step between the electron gun and the SCs. Other reasons were to control beam size and to suppress beam loss in and downstream of the SCs. At the end of the injector section, there is a profile monitor to observe the beam's shape and a Faraday cup to measure the amount of beam charge. An exhaust system is appropriately arranged to maintain ultra-high vacuum in this section. \par

Originally we used a 1.3 GHz buncher cavity identical to those at cERL. The beam velocity-modulated in the buncher cavity compresses the bunch length to about 30 ps at the first SC position (Z~2.4 m). This corresponds to a phase width of about $\pm$14 degrees for an acceleration frequency of 1.3 GHz. Bunch compression is seen at 1.3 GHz with RF phase nonlinear components and can also be a source of loss. It is possible to reduce these losses with a collimator, but it is difficult to compress the 100 ps bunch length with 1.3 GHz cERL buncher. Therefore, we redesign and apply the buncher cavity based on 650 MHz to increase the RF phase range and effectively perform bunch compression. \par

\begin{figure*}[!ht]
  \centering
  \includegraphics*[width=\textwidth]{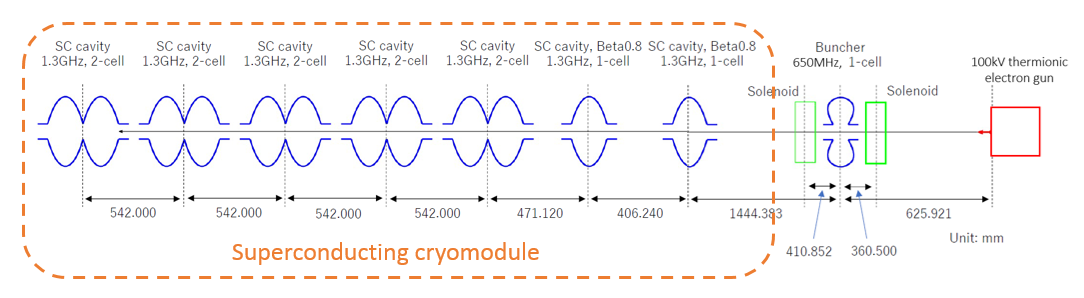}
  \caption{Layout of the injector and the superconducting cryomodule.}
  \label{fig:fig4}
\end{figure*}

\begin{figure}[!ht]
  \centering
  \includegraphics*[width=0.7\columnwidth]{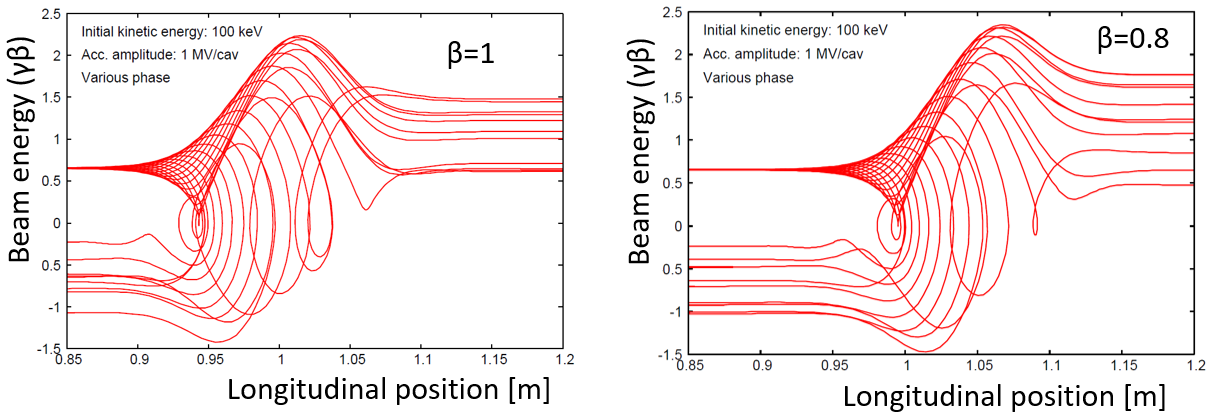}
  \caption{Minimum speed of transportation for $\beta=1$ cavity (left) and $\beta=0.8$ cavity (right).}
  \label{fig:fig3}
\end{figure}

The beam energy is accelerated up to 10 MeV by approximately 2 MV per cavity with five 2-cell cavities (Fig.\ref{fig:fig4}), which is feasible with Nb$_{3}$Sn. When accelerating in the 2-cell SC cavity after passing through the buncher, the acceleration does not occur immediately. This is because the energy of the beam (100 keV) is in the non-relativistic region, and the speed of the beam is not close to the speed of light. The 2-cell cavity does not accelerate properly due to the mismatch of the cavity cells, resulting in loss of energy. In order to eliminate cavity mismatch, each cell of the 2-cell cavities is made independent and the beam is slightly pre-accelerated by two 1-cell ($\beta=0.8$) cavities. As the graphs of the minimum speed of transportation (see Fig.\ref{fig:fig3}) read, if the minimum velocity is less than zero, the beam will decelerate, and it will be impossible to use it as an accelerator. If the amplitude of the cavity is increased too much, it will enter a region where the beam cannot be accelerated properly. The region that satisfies the condition that the energy finally rises without deceleration is wider for the $\beta=0.8$ cavity.

\section{Beam dynamics}
\label{sec:beam_dyn}

\subsection{Simulation setup}
The motivation of the simulation study was the following. Assuming an initial beam, optimization was performed by a simulation including the space charge effect of a multi-particle beam, so that the beam parameters at the exit of the accelerator were confirmed. 
\par

To study the beam dynamics in the compact Nb$_{3}$Sn accelerator, the injector part and the superconducting cryomodule were introduced into simulation. The electron gun distribution was included in this calculation, and the beam transport calculation was performed with the layout shown in Fig.\ref{fig:fig4}. We started the calculation by giving the particle distribution of the 100 keV beam at the cathode of the electron source. For the electron gun, solenoids, buncher and 2-cell cavities simulation the 1D field distributions of those from cERL were used. For the 1-cell cavity, the gap length of the original cavity was reduced ($\beta=0.8$), and the 1D field distribution was obtained. General Particle Tracer \cite{gpt} was used for the calculation. For the initial beam parameters please refer to Table~\ref{tab:table1}. \par

\begin{table}[!hbt]
 \caption{Initial parameters of electron beam}
  \centering
  \begin{tabular}{lcc}
    \toprule
           \textbf{Parameter} & \textbf{Value}  \\
       \midrule
    Bunch charge    & 77 pC (50 mA at 650 MHz) \\
    Number of particles & $10^6$ in tracking, \\
    { } & {2000 in optimization}    \\
    Electron gun energy     & 100 keV      \\
    Cathode size     & 5.73 mm (uniform)  \\
    Bunch length     & 100 ps (Gaussian $\pm4\sigma$)  \\
    \bottomrule
  \end{tabular}
  \label{tab:table1}
\end{table}

The electron emission surface of the hot cathode is approximately 10 mm, and the mean transverse energy (MTE) is 0.5 eV is used to generate a beam. The beam diameter at the electron gun exit is approximately 8 mm at its maximum. To control the space charge effect solenoid focusing is introduces into the design. Thus, the first solenoid focuses the beam, so that the beam diameter becomes 10 mm or less at the collimator position. Then the beam expanding on passing through the buncher cavity is focused again by the second solenoid. The beam size reaches the maximum diameter of 28 mm at the second solenoid, which is smaller than the beam pipe diameter of 60 mm. Therefore, it can be inferred that the beam loss at this location can be sufficiently suppressed. Another point is that the essential beam size relaxation at the second solenoid location allows in the following effectively shrink the transverse beam size in the cryomodule.\par

The bunch length primary estimated by the simulation was about 63 ps, but the experimentally obtained bunch length in Ref.\cite{stefani} is longer than the calculated bunch length. Thus, in a particle tracking an initial bunch length of 100 ps was adopted.  \par

\subsection{Optimization strategy}
The optimization strategy was: for a given layout of the accelerator, it was necessary to adjust the amplitudes and phases of cavities, and strengths of the solenoids to find the optimum conditions. Namely, to reach the target energy (10 MeV), the energy width of the emitted beam was required to be narrow, so that there was no loss during transportation.
For low velocity beams the parameters are intricately correlated. In addition, the influence of the space charge effect is also large. For this reason, simultaneous optimization of multiple parameters was necessary.\par

Optimization targets were the following: (1) to minimize final bunch length; (2) to minimize maximum beam size during transportation; (3) minimize final energy spread; (4) minimize maximum amplitudes of the cavities; (5) final energy should be reached. Then the distribution of 2,000 particles was set up, and the magnetic field of the solenoids on the transport of the beam, the acceleration voltages and the phases of the buncher and SCs were set as free parameters. The beam energy is accelerated up to 10 MeV by approximately 2 MV per cavity with five 2-cell cavities. The acceleration gradient is below 10 MV/m. Bayesian optimization  toolbox \cite{gpyopt2016} and solver based on multidimensional Newton-Raphson algorithm \cite{book} were used independently for the optimization to reach a reliable model of the accelerator. Their results converged. Then the acceleration voltages and phases of the buncher and SCs were confirmed. Together with it, the solenoids suppressed beam divergence due to the large initial space charge effect. As the final result of the optimizations, it was possible to transport 1,000,000 particles without hitting the 70 mm beam pipe.

\subsection{Results and discussion}
\begin{figure}[!ht]
  \centering
  \includegraphics[width=\columnwidth]{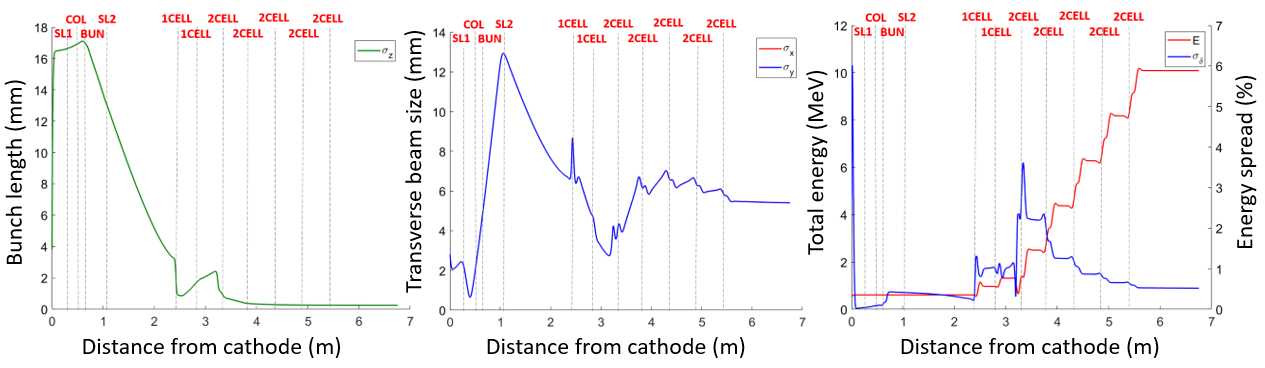}
  \caption{Time evolution of the beam parameters through the transportation line: rms bunch length (left); rms transverse beam size (middle); energy and energy spread (right).}
  \label{fig:fig5}
\end{figure}

The final results of the optimization procedure are summarized in Table~\ref{tab:table2}. Once all free parameters of the model were confirmed, the particle distribution was tracked through the transportation line. The evolution of the beam parameters is shown in Fig.\ref{fig:fig5}. And the beam performance at the exit of the cryomodule is given in Table~\ref{tab:table3}. \par

\begin{table}[!hbt]
 \caption{Optimized parameters}
  \centering
  \begin{tabular}{lccc}
    \toprule
    \textbf{Cavity name}     & \textbf{Amplitude}     & \textbf{Crest $\phi$}  & \textbf{$\phi$ off.}\\
    { }     & \textbf{MV/cav.}     & \textbf{deg.}  & \textbf{deg.}\\
    \midrule
    Buncher           & 0.0024  & -58.8 & -90.0     \\
    INJ1 (1-cell)     & 0.36 & -62.8 & 1.0      \\
    INJ2 (1-cell)     & 0.35 & -127.9 & 0.0      \\
    INJ3 (2-cell)     & 1.19 & 145.4 & -85.9     \\
    INJ4 (2-cell)     & 1.88 & -42.2 & 4.2      \\
    INJ5 (2-cell)     & 1.89 & -174.5 & 0.0      \\
    INJ6 (2-cell)     & 1.90 & 56.4 & 0.0      \\
    INJ7 (2-cell)     & 1.91 & 133.6 & 0.0      \\
    \midrule
    \textbf{Solenoid name}     & \textbf{Current (A)}     &    &  \\
    \midrule
    SL1     & 2.75     &    &  \\
    SL2     & 1.24  &    &  \\
    \bottomrule
  \end{tabular}
  \label{tab:table2}
\end{table}

Let's discuss the major knobs that could be used to control and to tune the beam. To control the bunch length effectively a combination of the buncher phase and the first 2-cell cavity phase offsets are used in the model. Buncher cavity involves so-called "zero-cross bunching" to compress the bunch and to leave its phase space linear. Then the bunching process launched at the buncher is stopped using a velocity bunching technique in the first 2-cell cavity (phase offset added to the crest phase). The bunch length behaviour is given in Fig.\ref{fig:fig5} to the left. Target compression achieved (see Table~\ref{tab:table3}). \par

\begin{figure}[!ht]
  \centering
  \includegraphics*[width=0.7\columnwidth]{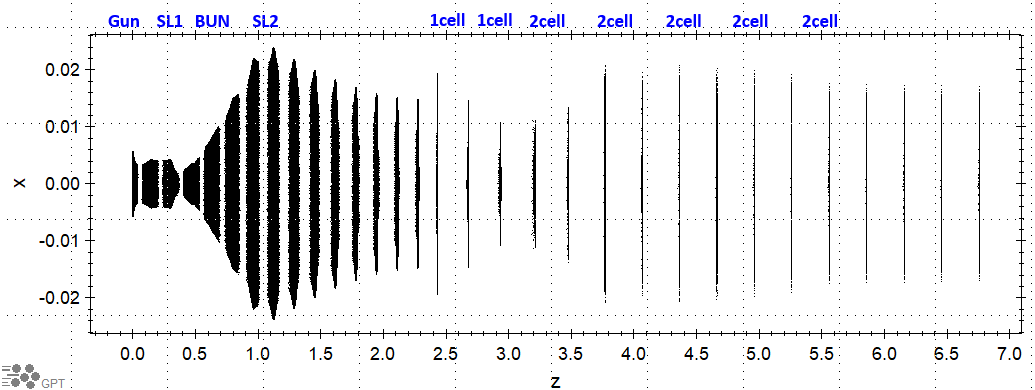}
  \caption{Time evolution of the spatial distribution of the beam.}
  \label{fig:fig6}
\end{figure}

\begin{figure}[!ht]
  \centering
  \includegraphics*[width=0.7\columnwidth]{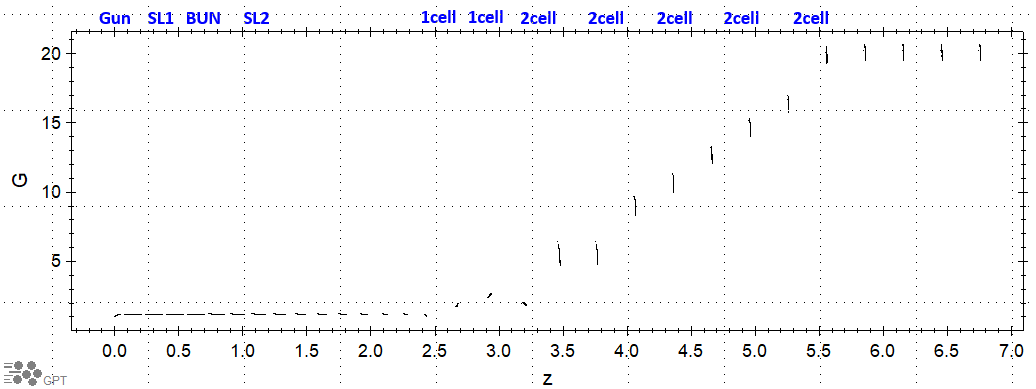}
  \caption{Time evolution of the energy distribution of the beam.}
  \label{fig:fig7}
\end{figure}

\begin{figure}[!ht]
  \centering
  \includegraphics*[width=0.7\columnwidth]{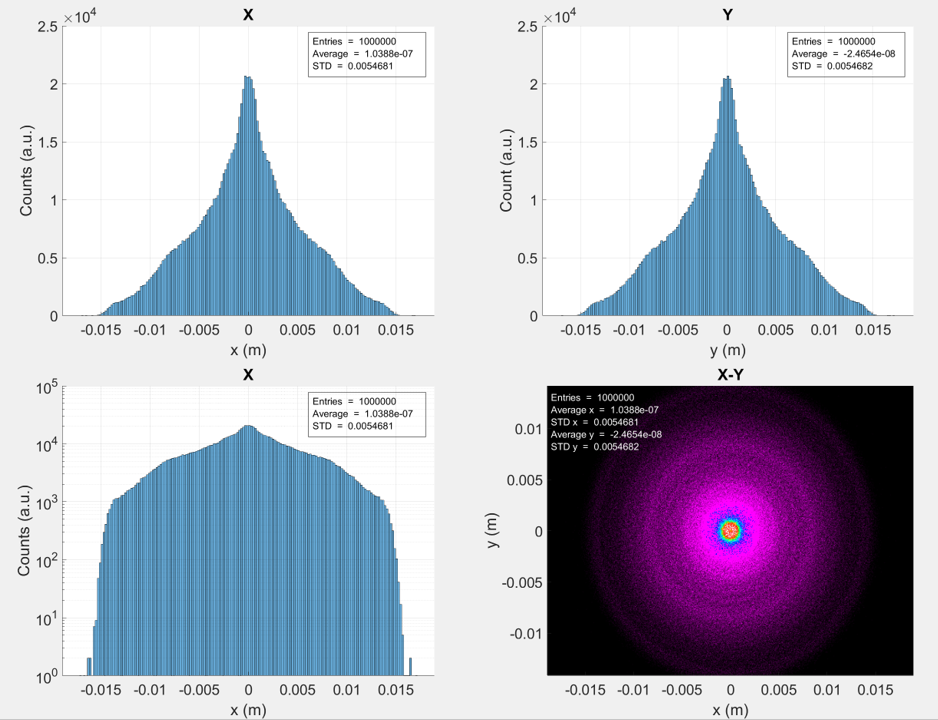}
  \caption{Final beam transverse spatial distribution 1m downstream the last cavity.}
  \label{fig:fig8}
\end{figure}

\begin{figure}[!ht]
  \centering
  \includegraphics*[width=0.7\columnwidth]{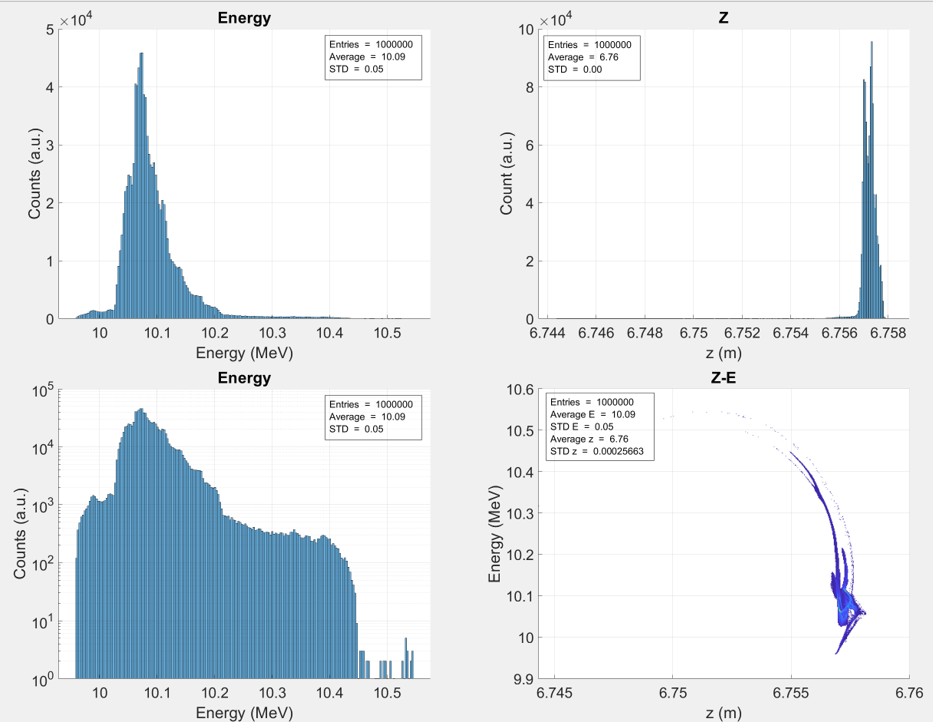}
  \caption{Final beam longitudinal spatial distribution 1m downstream the last cavity.}
  \label{fig:fig9}
\end{figure}

A key knob to control the transverse beam size is solenoid focusing (Fig.\ref{fig:fig5} in the middle). As it was mentioned above, the relaxation of the beam size at the location of the second solenoid (after the buncher cavity) is crucial for the minimization of the beam size in the presence of the strong space charge force. The cavity focusing effect turned to make a small impact into overall beam size minimization. Nevertheless, the beam size at the exit of the cryomodule is good (5.41 mm). Additional tuning available through 2 quadrupoles at the irradiation section.\par

Since the main acceleration occurs in five 2-cell cavities, energy tuning knobs are naturally amplitudes of those cavities. One should keep in mind, that the first 2-cell cavity is responsible for the velocity bunching. Therefore the fine energy tuning should be done through the rest four 2-cell cavities. Let us remind that 1-cell cavities are indispensable for the slight acceleration of the 100 keV beam coming from the electron source. This stage is of a great importance to reach the final energy of 10 MeV with a cryomodule. Another point is on the energy spread behaviour. It increases after accelerating in two 1-cell cavities since those fields' fringes oppose a considerable effect on the beam as reads the energy graph in Fig.\ref{fig:fig5} to the right.

Standalone optimization of injector parameters up to the second solenoid (see Fig.\ref{fig:fig4}) leads to the best beam parameters at the second solenoid location. However, with the optimization of beam parameters at the exit of the accelerating section, the same beam quality at the second solenoid location is not possible. The final result of the optimization of the beam parameters at the accelerator exit is achieved by relaxing those parameters at the second solenoid location. In this case, beam tracking occurs without losses.

Figure 6 shows a series of snapshots of the particle distribution in real space, taken every 1 nanosecond (ns). We observe the bunches being transported while converging in both the vertical and horizontal directions. This convergence is achieved by the buncher and solenoid focusing the beam. However, as the space charge effect becomes more prominent, the beam starts to diverge.  Following this divergence, the beam converges again due to acceleration.  The space charge effect also leads to significant scattering of some beam components, which can contribute to the formation of a beam halo and ultimately result in beam loss. These scattered components are evident in Figure 6.

Figure 7 presents overlapping snapshots of the particle distribution in longitudinal phase space, again captured every 1 ns. Here, we see the bunch being transported while converging in both energy and time.

Figures 8 and 9 depict the final particle distributions in the transverse and longitudinal planes, respectively. It is important to note that no beam loss was observed in the simulation with 1M macroparticles.

\begin{table}[!hbt]
 \caption{Beam parameters at the exit of the cryomodule}
  \centering
  \begin{tabular}{lc}
    \toprule
    Parameter & Value \\
    \midrule
    Total energy & 10.09 MeV \\
    Rms bunch length & 0.29 mm / 0.86 ps \\
    Rms transverse beam size & 5.41 mm \\
    Energy spread & 0.52\% \\
    \bottomrule
  \end{tabular}
  \label{tab:table3}
\end{table}

\section{SUMMARY}
Assuming an initial beam distribution, we performed three-dimensional beam tracking including the space charge effect. As a result of optimization by two independent methods mentioned above, the result of transportation with no loss was obtained. Based on the above calculation results and on the design of the injection section, Nb$_{3}$Sn cryomodule, and irradiation section \cite{sakai-IPAC23-THPM124}, we designed the entire irradiator. By using the Nb$_{3}$Sn cavities, it is possible to accelerate 10 MeV and 50 mA without loss, and a very compact irradiation accelerator can be achieved.

\section*{Acknowledgments}
This work is supported by the NEDO project “Development of innovative quantum beam technology for high-efficiency nanocellulose (CNF) production”.

%Bibliography
\bibliographystyle{unsrt}  
\bibliography{references}

\end{document}